\documentclass[sigconf]{acmart}

\usepackage{multicol}

\AtBeginDocument{%
  \providecommand\BibTeX{{%
    \normalfont B\kern-0.5em{\scshape i\kern-0.25em b}\kern-0.8em\TeX}}}

\copyrightyear{2022} 
\acmYear{2022} 
\setcopyright{acmcopyright}\acmConference[UbiComp/ISWC '22 Adjunct]{Proceedings of the 2022 ACM International Joint Conference on Pervasive and Ubiquitous Computing}{September 11--15, 2022}{Cambridge, United Kingdom}
\acmBooktitle{Proceedings of the 2022 ACM International Joint Conference on Pervasive and Ubiquitous Computing (UbiComp/ISWC '22 Adjunct), September 11--15, 2022, Cambridge, United Kingdom}
\acmPrice{15.00}
\acmDOI{10.1145/3544793.3563411}
\acmISBN{978-1-4503-9423-9/22/09}

\acmConference[UbiComp/ISWC '22 Adjunct]{Proceedings of the 2022 ACM International Joint Conference on Pervasive and Ubiquitous Computing}{September 11--15, 2022}{Cambridge, United Kingdom}

\acmBooktitle{Proc.\ of the 2022 ACM International Joint Conference on Pervasive and Ubiquitous Computing (UbiComp/ISWC '22 Adjunct)
}
\acmPrice{15.00}
\acmISBN{978-1-4503-9423-9/22/09}

\begin{document}

\title[Social Media App Usage \& Depression Scores during COVID-19 Pandemic]{Social Media App Usage in Relation with PHQ-9 Depression Scores during the COVID-19 Pandemic}

\author{Lena Mulansky, R{\"u}diger Pryss}
\affiliation{%
  \institution{Institute of Clinical Epidemiology and Biometry, University of W{\"u}rzburg}
  \streetaddress{Josef-Schneider-Str. 2}
  \city{W{\"u}rzburg}
  \country{Germany}
  \postcode{97080}
}
\email{lena.mulansky@uni-wuerzburg.de}
\email{ruediger.pryss@uni-wuerzburg.de}

\author{Caroline Cohrdes}
\affiliation{%
  \institution{Mental Health Research Unit, Department of Epidemiology and Health Monitoring, Robert Koch Institute}
  \city{Berlin}
  \country{Germany}}
\email{CohrdesC@rki.de}

\author{Harald Baumeister}
\affiliation{%
  \institution{Department of Clinical Psychology and Psychotherapy, Ulm University}
  \city{Ulm}
  \country{Germany}
  \postcode{89081}}
\email{harald.baumeister@uni-ulm.de}

\author{Felix Beierle}
\affiliation{%
  \institution{National Institute of Informatics}
  \city{Tokyo}
  \country{Japan}
  }
\affiliation{%
  \institution{Institute of Clinical Epidemiology and Biometry, University of W{\"u}rzburg}
  \city{W\"urzburg}
  \country{Germany}
  }
\email{beierle@nii.ac.jp}

\renewcommand{\shortauthors}{Mulansky et al.}

\begin{abstract}
With about 300 million affected people, major depressive disorder (MDD) is one of the most common diseases worldwide. During the COVID-19 pandemic, the number of cases increased even further, by 28\%. Many factors may be correlated with MDD, including the excessive use of social media apps. In this paper, we investigated the relationship between the use of social media and communication apps and depressive symptoms during the COVID-19 pandemic. The pandemic and social distancing like lockdowns probably changed smartphone usage times and usage patterns. While previous studies have shown an association between depression and social media usage, we report about the situation during these special circumstances. We employed a log-linear regression to examine the association of social media and communication app usage and depression. To quantify the usage, we applied the total usage time in hours of social media apps (e.g., WhatsApp, Facebook) as well as communication apps (Phone and Messaging) within one week. To measure depressive symptoms, we used the PHQ-9 score. We discovered a significant association between the usage time and the PHQ-9 score (beta=0.0084, p-value=0.010). We conclude that social media usage is a robust marker for depression severity and future research should focus on a better understanding of the underlying causality and potential counter-measures.
\end{abstract}

\ccsdesc[500]{Applied computing~Psychology}
\ccsdesc[500]{Applied computing~Health informatics}

\keywords{COVID-19; social media; app usage; depression}

\maketitle

\section{Introduction}
Major depressive disorder (MDD) is one of the leading mental health diseases in the world. Approximately 300 million people worldwide suffer from depressive disorder \cite{whoprevalence2017}.
During the COVID-19 pandemic, MDD has increased even further. \cite{santomauroGlobalPrevalenceBurden2021} estimated a global increase of about 28\% due to the COVID-19 pandemic.
The subsequent costs of MDD are enormous.
In the US, MDD leads to a total cost of approximately 326 billion US-Dollar per year.
\cite{greenbergEconomicBurdenAdults2021}. Despite the increasing awareness of MDD, the disease is often diagnosed late or not at all and therefore often remains untreated \cite{lotfiUndetectedDepressionPrimary2010,mitchellClinicalDiagnosisDepression2009}. Early diagnosis of MDD would significantly improve the chances of treatment and thus also reduce the high costs incurred by the disease \cite{habertFunctionalRecoveryMajor2016,greenbergEconomicBurdenAdults2021}. There are multiple factors that significantly 
correlate with
MDD. In existing studies, the effects of many demographic and non-demographic factors, such as age \cite{stordalAssociationAgeDepression2003}, gender \cite{salkGenderDifferencesDepression2017} or the household size \cite{sempunguChangesHouseholdSize2022} on MDD have already been examined.
Also, the correlation of  depressive symptoms with smartphone usage \cite{elhaiFearMissingOut2016} and especially with social media usage have been investigated in several studies in the past (e.g., \cite{PassiveActiveSocial,robinsonSocialComparisonsSocial2019,moshePredictingSymptomsDepression2021}).
During the COVID-19 pandemic lockdown, social media took on a different meaning. People used social media apps to cope with feelings of loneliness and anxiety due to the lack of physical contact, and also as a source of information \cite{HowAdolescentsUse}.
In this study, we examine the relationship between social media and communication app usage and depressive symptoms during the pandemic. The association of social media apps and MDD has already been investigated, but in most of the studies, the researchers had to rely on self-reported smartphone usage time information and could not assess mobile sensing data. Through mobile sensing, it is possible to acquire additional people-centered or environment-centered data using the sensors of the mobile devices. Measuring the usage of communication apps via mobile sensing, after the permission of the user is given, represents a more objective way of assessing the usage time. In this study, mobile sensing data in form of app usage time is applied. 
The mental health information of the users is gathered through an Ecological Momentary Assessment (EMA) questionnaire. EMA is used to assess momentary feelings, behavior or symptoms, typically via self-reports. Our main contribution is to reveal the association of the usage time of social media and communication apps with depressive symptoms during the COVID-19 pandemic.
These associations can help researchers and clinicians to understand to what extent the pandemic might have influenced overall
social media usage
and if pre-pandemic research results about the links with depression still hold true.

\section{Related Work}
\label{sec:relwork}

Lin et al.\ \cite{linAssociationSocialMedia2016} as well as Escobar-Viera et al.\ \cite{PassiveActiveSocial} examined the relationship between social media use and depression. In \cite{PassiveActiveSocial}, the authors analyzed whether there is a difference between the impacts of active and passive social media use on depression.
In \cite{robinsonSocialComparisonsSocial2019}, they looked at the relationship between social media use and MDD as well, however, in contrast to \cite{linAssociationSocialMedia2016} and \cite{PassiveActiveSocial}, they focused their investigation on specific social media behaviors within a group of millennials.
\cite{linAssociationSocialMedia2016} and \cite{robinsonSocialComparisonsSocial2019} found out that people with a higher total use time are significantly more likely to meet the criteria of MDD compared to those with a lower total use time. The authors of \cite{PassiveActiveSocial} found out that an increased passive use of social media apps is positively correlated with more depressive symptoms.

Rozgonjuk et al.\ \cite{rozgonjukAssociationProblematicSmartphone2018} used smartphone daily screen time and phone screen unlocks to explore its relation to depression. They found that depression is not related to screen time minutes, but is negatively correlated with the frequency of phone screen unlocking.
\cite{elhaiNonsocialFeaturesSmartphone2017}
highlighted the differences of the relationship between social smartphone usage and process usage and depressive symptoms. They detected a negative association between depression symptom severity and greater social smartphone use. Wetzel et al.\ have investigated the informative effect of smartphone usage data on loneliness and social well-being \cite{wetzelHowComeYou2021}.
They found a negative relationship between younger participants with higher use times and social well-being and a positive association with loneliness, while the opposite association was found for older adults. In
\cite{moshePredictingSymptomsDepression2021}, smartphone usage data was used to examine, whether those could predict symptoms of depression. The authors focused on GPS data and smartphone total usage rather than social media use time and found a significant negative association between the variability of locations visited and symptoms of depression. In \cite{linAssociationSocialMedia2016,PassiveActiveSocial,robinsonSocialComparisonsSocial2019,elhaiNonsocialFeaturesSmartphone2017} the usage data was self-reported by the users, whereas in \cite{rozgonjukAssociationProblematicSmartphone2018,wetzelHowComeYou2021,moshePredictingSymptomsDepression2021}, they are based on mobile sensing app data.
Moreover, \cite{linAssociationSocialMedia2016,robinsonSocialComparisonsSocial2019,rozgonjukAssociationProblematicSmartphone2018} had only young adults or students included in their investigation, whereas \cite{PassiveActiveSocial,elhaiNonsocialFeaturesSmartphone2017,wetzelHowComeYou2021,moshePredictingSymptomsDepression2021} contain a larger age range. To measure the level of depression, the authors used different scales and method: \cite{linAssociationSocialMedia2016,PassiveActiveSocial} used the Patient-Reported Outcomes Measurement Information System (PROMIS) Depression Scale Short Form, \cite{rozgonjukAssociationProblematicSmartphone2018,elhaiNonsocialFeaturesSmartphone2017,moshePredictingSymptomsDepression2021} used the Depression Anxiety Stress Scales (DASS-21) and \cite{robinsonSocialComparisonsSocial2019} the PHQ-9 scale.

Some of the cited studies are based on self-reported details about the social media use time of users, and all investigations were conducted before the COVID-19 pandemic. 
With the possibly significant changes in smartphone usage because of the pandemic and social distancing measures, we investigate the association between social media usage and depression during the COVID-19 pandemic.
We use a large sample (n=627) and employ data from
mobile sensing, allowing for objective measurements.

\section{Methodology}
\label{sec:methods}

The data used in this study were gathered through the Corona Health app \cite{beierleCoronaHealthStudy2021,holfelderMedicalDeviceRegulation2021}.
The participants completed the questionnaire voluntarily and could also give permission to access app usage history. This mobile sensing data could only be collected for Android users, because it was not available to access for software developers on iOS devices. The mobile sensing data comprise the daily use time of common smartphone communication apps and were gathered for 7 days prior to submitting the questionnaire~\cite{beierleCoronaHealthStudy2021}.

\begin{table}[ht]
\footnotesize
\centering
\caption{Data Characteristics (SD = Standard Deviation).}
\label{tab:characteristics}
\begin{tabular}{lrrrrrrrrr}
\toprule
{\textbf{Variable}} & \textbf{n(\%)} & \textbf{Min} & \textbf{Max} &\textbf{Mean} &\textbf{SD} \\ \hline
Age & 627 & 18 & 78 & 41.97 & 13.59\\
\hspace{2mm} 18-29 & 128 (20.4\%) & 18 & 29 & 23.70 & 3.52 \\
\hspace{2mm} 30-44 & 238 (37.9\%) & 30 & 44 & 36.93 & 4.07 \\
\hspace{2mm} 45-59 & 185 (29.5\%) & 45 & 59 & 51.83 & 4.30 \\
\hspace{2mm} 60-78 & 76 (12.1\%) & 60 & 78 & 64.50 & 4.35 \\
Gender & 627 & - & - & - & - \\
\hspace{2mm} Male & 296 (47.2\%) & - & - & - & -\\
\hspace{2mm} Female & 331 (52.8\%) & - & - & - & - \\
Use time (hours/week) &  &  \\
\hspace{2mm} $<$ 5 & 342 (54.5\%) & 0.00 & 4.97 & 1.96 & 1.52 \\
\hspace{2mm} 5-10 & 127 (20.3\%) & 5.01 & 9.99 & 7.29 & 1.44\\
\hspace{2mm} 10-15 & 82 (13.1\%) & 10.03 & 14.82 & 12.16 & 1.36 \\
\hspace{2mm} 15-20 & 36 (5.7\%) & 15.08 & 19.98 & 17.47 & 1.57 \\
\hspace{2mm} $>$ 20 & 40 (6.4\%) & 20.17 & 59.12 & 29.26 & 9.01 \\
PHQ-9 score & 627 & 0 & 27 & 9.34 & 6.57  \\
\hspace{2mm} 0-4 & 176  (28.1\%) & 0 & 4 & 2.17 & 1.35 \\
\hspace{2mm} 5-9 & 186 (29.7\%) & 5 & 9 & 6.95 & 1.42 \\
\hspace{2mm} 10-14 & 127 (20.3\%) & 10 & 14 & 11.94 & 1.45 \\
\hspace{2mm} 15-19 & 78 (12.4\%) & 15 & 19 & 16.96 & 1.47 \\
\hspace{2mm} 20-27 & 60 (9.6\%) & 20 & 27 & 22.35 & 1.96 \\
\bottomrule
\end{tabular}
\end{table}

Besides demographic information, the data of the questionnaire include also psycho-social health information of the users.
The cross-sectional data
is from
July 2020 to June 2021 and, after preprocessing, includes 627 users between 18 and 78 years.
To indicate the depression level, the depression score was calculated from self-reported answers of the nine questions of the PHQ-9 questionnaire.
The score ranges from 0 to 27, indicating the severity of the depression
(0-4 minimal, 5-9 mild, 10-14 moderate, 15-19 moderately severe,
20-27 severe) \cite{kroenkePHQ92001}.
The use of social media and communication apps was measured by the complete active use time in hours in the past week of the following apps: Facebook and Facebook Messenger, Instagram, Messaging, Phone, Skype Video, Snapchat, Telegram, and WhatsApp.
Time in which the app was open in the background was not included in the analysis.
During the preprocessing, users who did not give the permission to collect mobile sensing data, have been dropped.
To examine whether the use of social media and communication apps is associated with depression, we conducted a regression with the PHQ-9 score as dependent variable and the overall use time in hours of the above-mentioned apps during the past week as explanatory feature. As control variables, we included age and gender (as a binary variable), motivated by existing significant results \cite{stordalAssociationAgeDepression2003,salkGenderDifferencesDepression2017,whoprevalence2017}.
We had to exclude a small number of users of diverse sex (n=6) from the analysis because each value should occur at least 10 times in the used data set \cite{harrell1996multivariable}.
Prior to the statistical analysis, we logarithmized the dependent variable, since the PHQ-9 score was skewed to the right and therefore the normality assumption of the residuals was not given. So we used a log-linear regression model. All other preconditions for a regression are given and verified.

As presented in Table \ref{tab:characteristics}, the final sample includes 627 participants with 52.8\% females and 47.2\% males. Most participants were aged between 30 and 44 years (38\%). Almost 55\% of all participants had a total use time of social media and communication apps less than 5 hours in one week. About 12\%  used them 15 or more hours in one week.
Over 20\% of all participants have moderately severe or severe depressive symptoms (PHQ-9 score equals 15 or higher). Table \ref{tab:use time} shows the different distribution of the total app use time and the PHQ-9 score within the different age groups and between male and female users. The mean total use time of participants between 18 and 29 years is 10.41 hours per week, and thus larger than in the other three age groups. The average PHQ-9 score in this age group is 12.18, which is also the highest mean value of all groups. Moreoever, it can be seen that the average total use time as well as the mean PHQ-9 score of female participants is higher than the average use time and score of male users.

\section{Results}
\label{sec:results}

\begin{table*}[ht]
\footnotesize
\centering
\caption{Distribution of Use Time and PHQ-9 Score.}
\label{tab:use time}
\setlength{\tabcolsep}{9pt} 
\begin{tabular}{lrrrrrrrr}
\toprule
{\textbf{  }} & \multicolumn{4}{c}{\textbf{Total use time}}  & \multicolumn{4}{c}{\textbf{PHQ-9 score}}\\
\hline
{\textbf{Variable}} & \hspace{3mm} {\textbf{Min}} & {\textbf{Max}} & \textbf{Mean} & \textbf{SD} & \hspace{3mm} {\textbf{Min}} & {\textbf{Max}} &\textbf{Mean}&\textbf{SD}\\
\hline 
Age &  &  \\
\hspace{2mm} 18-29 & \hspace{3mm} 0.04 & 52.25 & 10.41 & 9.37 & \hspace{3mm} 0 & 26 & 12.18 & 7.00\\
\hspace{2mm} 30-44 & \hspace{3mm} 0.00 & 59.02 & 6.91 & 7.84 & \hspace{3mm} 0 & 27 & 9.54 & 6.39\\
\hspace{2mm} 45-59 & \hspace{3mm} 0.00 & 42.71 & 5.90 & 7.00 & \hspace{3mm} 0 & 23 & 8.10 & 5.94\\
\hspace{2mm} 60-78 & \hspace{3mm} 0.00 & 27.14 & 4.27 & 5.28 & \hspace{3mm} 0 & 25 & 6.93 & 6.19\\
Gender &  &  &  &   \\
\hspace{2mm} Male & \hspace{3mm} 0.00 & 59.12 & 5.71 & 7.15 & \hspace{3mm} 0 & 26 & 8.58 & 6.73\\
\hspace{2mm} Female & \hspace{3mm} 0.00 & 52.25 & 8.17 & 8.36 & \hspace{3mm} 0 & 27 & 10.02 & 6.36\\
\bottomrule
\end{tabular}
\end{table*}

We present the log-linear model in Table \ref{tab:regression}. The overall use time of
social media and communication apps is significantly related to the PHQ-9 score ($beta=0.0084$, $p<0.05$), however, it does not add much additional explanatory value to the indicators of a depression. If a user spends one hour more for using the communication apps,
the depression score is about 0.84\% higher. As expected, also age and gender have a significant impact on the score (both $p<0.05$). The PHQ-9 score of a male user is on average 
about 16.4\% lower than the score of a female user, whereas the score decreases about 0.96\% on average per year of life.

\section{Discussion}
\label{sec:discussion}

The results of the log-linear regression are significant ($p=0.010$) and suggest a negative association of higher social media and communication app usage and depressive symptoms ($beta=0.0084$).
Former studies have shown similar results before the COVID-19 pandemic.
During the pandemic, social distancing measures
like lockdowns might have altered the overall smartphone and social media usage pattern, creating the need to study
if the association between social media use and depression still exists.
The results of our study highlights that this established link still holds true during the COVID-19 pandemic.
Thus, we need to aim at a better understanding on the causality of this association, as social media app usage might be a dysfunctional coping strategy for depression and/or depression severity might be negatively influences by social media app usage.
The greatest strength of our study is the large sample size (n=627) compared to similar studies in which mobile sensing data was used (e.g., n=101 \cite{rozgonjukAssociationProblematicSmartphone2018} and n=55 \cite{moshePredictingSymptomsDepression2021}). Moreover, our sample has a much larger age range (18-78 years) than other studies \cite{linAssociationSocialMedia2016, robinsonSocialComparisonsSocial2019}.
Additionally, we used objective app usage time measurements via mobile sensing rather than self-reported usage data to avoid a potential self-report bias \cite{althubaitiInformationBiasHealth2016}.

\begin{table*}[ht]
\footnotesize
\centering
\caption{Log-Linear Regression Results (CI = Confidence Interval).}
\label{tab:regression}
\setlength{\tabcolsep}{9pt} 
\begin{tabular}{lrrrrl}
\toprule
{\textbf{Predictors }} & \textbf{Beta} & \textbf{Std Error} & \textbf{t-value} & \textbf{p-value} & \textbf{[95\% CI]}\\ 
\hline 
Intercept & 2.3549 & 0.092  & 25.488 & 0.000 & [2.173;2.536]\\
Age & -0.0096 & 0.002 & -5.158 & 0.000 & [-0.013;-0.006]\\
Male & -0.1639 & 0.050  & -3.299 & 0.001 & [-0.262;-0.066]\\
Total use time & 0.0084 & 0.003  & 2.594 & 0.010 & [0.002;0.015]\\
\hline
 {\textbf{  }} & \textbf{Adj. R-squared} & \textbf{0.083} & \textbf{} & \textbf{}\\ 
 \bottomrule
\end{tabular}
\end{table*}

Through the COVID-19 pandemic and the resulting tendency towards work from home, some social media and communication apps might have been used for work.
In the variable 'use time' we could not differentiate between private and work use.
In \cite{linAssociationSocialMedia2016}, they explicitly asked about the personal usage of social media apps.
In former studies, the authors mainly focus on social media usage \cite{linAssociationSocialMedia2016,PassiveActiveSocial,robinsonSocialComparisonsSocial2019}, while our study encompasses also the usage time of phone calls and messaging.
The boundary between what constitutes a messaging app and what constitutes a social media app are likely fluid.
Note that past studies might have tracked other apps because of changes in app popularity.
For example, \cite{linAssociationSocialMedia2016} included Tumblr, Twitter, Google+, Youtube, LinkedIn, Pinterest, Vine and Reddit besides Facebook and Instagram, whereas we considered the usage time of Facebook, Facebook Messaging, Instagram, Messaging, Phone, Skype Video, Snapchat, Telegram, and WhatsApp.
When interpreting the presented results, some limitations have to be considered. We could only include Android users, therefore a selection bias cannot be ruled out.
The sample only contains German users over 17 years, and, additionally, age is not distributed equally across the sample. Most of the people were between 30 and 44 years old.
Another limitation might be that
we only considered a limited number of the most frequently used communication apps and not all of them.
However, as the most frequently used apps are included, we do not expect the results to change by much when including smaller social media and communication apps.
Furthermore, the usage time of social media was only measured via smartphone usage, however, people can also use platforms like Facebook on their computer.

\section{Conclusion}
\label{sec:conclusion}

In this paper, we analyzed the relation between social media and communication app use and depression, measured by PHQ-9, during the COVID-19 pandemic.
Our statistical analysis shows a significant association, showing higher depression scores
for higher app use.
While existing studies have shown such relations, to the best of our knowledge,
we are the first to report that these
results hold true even during
social distancing measure like lockdowns,
during which we expect overall increased
app use.
Hence, social media app usage seems to represent a robust marker for depression severity, asking for a better understanding of the underlying causality and likely for counter-measures in order to maintain public mental health in times of excessive smartphone usage.

\begin{acks}
This work was supported by a fellowship within
the IFI program of the German Academic Exchange Service (DAAD).
\end{acks}

\bibliographystyle{ACM-Reference-Format}

\end{document}